\documentclass[prd,showpacs,twocolumn,nofootinbib,superscriptaddress]{revtex4}
\usepackage{graphicx} 
\usepackage{amsmath} 
\usepackage{amssymb}
\usepackage{subfigure}
\usepackage{psfrag}
\usepackage{slashed}

\newcommand{\intd}{{\textrm{d}}}

\newcommand{\vecc}[1]{{\vec{#1}}}
\newcommand{\intddd}[1]{\int \! \frac{\intd^3 #1}{(2\pi)^3}}

\newcommand{\beq}{\begin{equation}}
\newcommand{\eeq}{\end{equation}}

\newcommand{\MeV}{{\textrm{MeV}}}
\newcommand{\GeV}{{\textrm{GeV}}}

\newcommand{\millibarn}{{\textrm{mb}}}

\newcommand{\dorria}{r}

\begin{document}

\title{
Chiral restoration effects on the shear viscosity of a pion gas
}

\author{Klaus Heckmann}
\affiliation{Institut f\"ur Kernphysik, Technische Universit\"at Darmstadt, Germany}

\author{Michael~Buballa}
\affiliation{Institut f\"ur Kernphysik, Technische Universit\"at Darmstadt, Germany}

\author{Jochen~Wambach}
\affiliation{Institut f\"ur Kernphysik, Technische Universit\"at Darmstadt, Germany}
\affiliation{GSI Helmholtz-Zentrum f\"ur Schwerionenforschung, Darmstadt, Germany}

\begin{abstract}
\noindent
We investigate the shear viscosity of a pion gas in
  relativistic kinetic theory, using the Nambu-Jona-Lasinio model
  to construct the pion mass and the $\pi{}\pi$-interaction at finite
  temperature. Whereas at low temperatures the scattering
  properties and, hence, the viscosity are in agreement with lowest-order
  chiral perturbation theory, we find strong medium modifications in
  the crossover region. Here the system is strongly coupled and the
  scattering lengths diverge, similarly as for ultra-cold Fermi gases at 
  a Feshbach resonance. As a consequence, the ratio $\eta/s$ is found to 
  be strongly reduced as compared to calculations without medium-modified
  masses and scattering amplitudes. However, the quantitative results
  are very sensitive to the details of the applied approximations.
\end{abstract}


\date{October 2012}

\maketitle


\section{Introduction}

The elliptic flow at RHIC was found to be in rather good agreement
with almost ideal hydrodynamic behavior \cite{Huovinen-et-al:PLB01,Teaney-et-al:NPA02,Adare-et-al:PRL07,Adams-et-al:PRC05,RomatschkeLuzum:PRC08,RomatschkeLuzum:PRC09-Erratum,SongHeinz:JPG09}. This led to an intensive discussion of the
hydrodynamical properties of strong-interaction (QCD) matter under extreme conditions. 
The viscous properties of a thermal system are characterized by its transport coefficients. 
These describe the deviation from ideal hydrodynamics and are
determined by the micro-physics.

While originally the RHIC data have been interpreted as evidence for
an almost perfect fluid in the quark-gluon phase, it was pointed out
recently that the elliptic flow crucially depends on the transport
properties in the hadronic phase~\cite{Niemi-etal:arXiv2011}.
In this phase, at low temperatures and small baryo-chemical potentials,
the thermal properties of QCD matter are governed by the lightest degrees
of freedom, namely pions. Moreover, for small enough temperatures,
a gas of pions is sufficiently dilute to apply relativistic
kinetic theory. Such a kinetic description of transport processes
requires solely the (in-medium) masses and scattering amplitudes of the
underlying particles. In this context, usually the vacuum values of these
quantities are used~\cite{Prakash_etal:1993,DobadoPelaez:PRD93,DobadoPelaez:PRD98,Dobado_Santalla:01,DLlE:04,FernandezFraila:EPJA07,Chen:2006iga,Chen:2007xe,IOM:08}. 
The temperature dependence of the transport coefficients is then the result 
of the different thermal occupation of states. However, the pion mass and
the pion interactions are modified by the hot and dense medium. Since
pions are (near) Goldstone-bosons of spontaneously broken chiral symmetry, 
their properties are sensitive to its restoration at high temperatures, especially 
in the chiral crossover region.

A consistent description of pion scattering processes in the vacuum and at low
temperatures can be obtained within chiral perturbation theory
\cite{Lehmann:PLB72,Pagels:PR75,Weinberg:Physica79,GasserLeutwyler:AnnPhys84,Gerber:1988tt,Gerber:1990yb}.
However, the expansion breaks down in the chiral crossover region where
nonperturbative effects become essential.
Therefore, in order to describe $\pi{}\pi$-scattering in this region, an
approach is needed which allows for a proper description
of the chiral phase transition.
The starting point for our investigations will be the Nambu-Jona-Lasinio (NJL)
model~\cite{NJL:I,NJL:II}, which has been used extensively to study
spontaneous chiral symmetry breaking in the vacuum and its restoration at 
finite temperature and (net) quark density.
(For reviews see~\cite{VW91,Klevansky:Review92,HK94,Buballa:Review05}.)
Unlike in the linear sigma model, which was employed in ref.~\cite{DLlETR:09}
in a similar context, the elementary degrees of freedom in the NJL model
are quarks. Mesons are constructed from $q\bar{q}$-correlations.
At low temperatures, the pion emerges as a bound state in the pseudoscalar
channel, whose mass, decay constant and scattering properties are consistent
with chiral low-energy theorems~\cite{Schulze:JPG95,QuackKlevansky-etal:PLB95}.
However, in the vicinity of chiral restoration, the pion gets dissolved
and becomes a broad resonance at high temperatures.
This feature, which is a natural consequence of the compositeness, 
cannot be described properly in the linear sigma model and is our
main motivation to use the NJL model.

In the present work, we focus on the shear viscosity $\eta$. This transport
coefficient is essential for the behavior of the elliptic
flow in non-central heavy-ion collisions. Moreover, the 
ratio of shear viscosity over entropy density, $\eta/s$, is a measure for
the proximity of a given system to ideal fluid dynamical behavior.
In the context of the AdS/CFT correspondence it has been conjectured that  
$\eta/s$ has a universal lower bound of $1/4\pi$ \cite{KSS:PRL05}
(although this conjecture has been questioned by some authors, see, e.g.,
Ref.~ \cite{Cohen:2007qr}).
Thus, an 
almost-ideal fluid should acquire a value of $\eta/s$ which is close to this 
putative 
lower bound. Besides the hot and dense fireball of a heavy-ion collision, nearly-ideal 
hydrodynamic behavior has also been discussed for a system of ultra-cold fermionic 
atoms in a trap. Here a divergent scattering length (unitary limit) can be achieved 
via a Feshbach resonance tuned with an external magnetic field. Ultra-cold Fermi 
gases thus offer the exciting possibility to study strongly coupled systems under 
'controlled conditions' in the laboratory and infer analogies to QCD matter. We 
shall see that a divergent scattering length can also be found in a pion gas near 
the chiral crossover.

This article is organized as follows. In sect.~\ref{sec:short-rkt}, we
describe the main steps for calculating the shear viscosity in
relativistic kinetic theory.
In sect.~\ref{sec:njl}, we employ the NJL model to determine the relevant 
quantities which enter the kinetic description: After briefly reviewing 
the temperature dependence of quark and meson masses as well as   
the $\pi{}\pi$-scattering lengths, we introduce several approximation steps 
of increasing sophistication to calculate the in-medium $\pi{}\pi$-scattering 
amplitude and discuss the resulting changes of the total cross sections.  
The results for the shear viscosity and for $\eta/s$ are
presented in sect.~\ref{sec:results}. Finally, in sect.~\ref{sec:discussion},
we discuss the implications of our results and propose further developments.

\section{Shear viscosity from kinetic theory}
\label{sec:short-rkt}

In this section we summarize the main ideas of how to calculate the 
shear viscosity within kinetic theory and list the corresponding equations 
that need to be solved. We largely follow Refs.~\cite{DLlE:04,Chen:2006iga,Chen:2007xe,IOM:08}, 
and refer to these articles for further details.

The basic ingredients of relativistic hydrodynamics are the 
local fluid four-velocity $u^\mu(x)$ and the energy-moment\-um tensor 
$T^{\mu \nu}$. Performing an expansion in gradients of $u$, 
the latter can be written as
\beq
T^{\mu \nu} = T^{(0) \mu \nu} + T^{(1) \mu \nu} + \dots\,
\label{eq:T_munu}
\eeq
where
\beq 
T^{(0) \mu \nu} = (\epsilon + p)\,u^\mu u^\nu - p\,g^{\mu\nu} 
\label{eq:T_munu0}
\eeq
is the ideal part solely specified by the local energy density $\epsilon$ and pressure $p$,
while
\begin{align} 
T^{(1) \mu \nu} \;=\; 
&\eta \left[\partial^\mu u^\nu  + \partial^\nu u^\mu -
           u^\mu u^\lambda \partial_\lambda u^\nu -
           u^\nu u^\lambda \partial_\lambda u^\mu \right]
\nonumber \\
+ & \left( \zeta - \frac{2}{3}\eta \right)\,
     \left[g^{\mu\nu} - u^\mu u^\nu\right]\partial_\lambda u^\lambda
\label{eq:T_munu1}
\end{align}
gives the first-order viscous correction
with the bulk viscosity $\zeta$ and the shear viscosity $\eta$.

In order to determine these transport coefficients 
in kinetic theory one exploits the fact that $T^{\mu \nu}$ is related
to the phase-space distribution functions 
$f(x,p) \equiv f(\mathbf{x},\mathbf{p},t)$
of the particles involved.
In the present study we limit ourselves to a gas of pions. 
Then we have 
\beq
    T^{\mu \nu}(x) = 
    g_\pi \int \! \frac{\intd^3 p}{(2\pi)^3 E}\, p^\mu p^\nu 
    f(x,p)\,,
\label{eq:Tmunu_fpi}
\eeq
where $g_\pi=3$ is the isospin degeneracy of the pions and
$p^0 \equiv E = \sqrt{m_\pi^2 + \mathbf{p}^2}$ is their on-shell energy. 
In a relativistic quantum mechanical framework, the 
space-time evolution of the distribution function
is governed by the Boltz\-mann-Ueh\-ling-Uhlen\-beck
(BUU) equation
\beq
  \label{eq:short-BUU}
   \frac{p^\mu}{E} \partial_\mu f(x,p) 
   = C_{\pi \pi}[f],
\eeq
where the collision term $C_{\pi \pi}[f]$ encodes the changes
of the distribution function due to interaction processes.
In the following we restrict ourselves to $2 \to 2$ processes,
$p + p_1  \leftrightarrow  p' + p_1'$.
Then the collision term is given by
\begin{align}
  \label{eq:short-BUU-22}
  C_{\pi \pi}[f_\pi]
  = & \frac{g_\pi}{2}
  \intddd{p'}\intddd{p_1}  \intddd{p_1'}\,
  \times \nonumber \\ &  \times
  \Bigg\{
  |{\cal{M}}_{\pi \pi}|^2
  \frac{(2\pi)^4\delta^4(p+p_1-p'-p_1')}{16EE_1E'E_1'}
  \times \nonumber \\ &  \quad \times
  \big[f_{p'}f_{p_1'}(1+f_p)(1+f_{p_1})
  \nonumber \\ &  \qquad
  - f_p f_{p_1}(1+f_{p'})(1+f_{p_1'}) \big]
  \Bigg\},
\end{align}
where we have introduced the short-hand notation $f_p \equiv f(x,p)$.
The main physics input is the square of the isospin averaged invariant matrix element for $\pi{}\pi$-scattering:
\beq
\label{eq:Mpipi}
    |{\cal{M}}_{\pi \pi}|^2
    \equiv
    \frac{1}{9}  \sum_{I=0}^{2}\, (2I+1)
    \left| {\cal{M}}_{\pi{}\pi{}}^I \right|^2\,,
\eeq
where ${\cal{M}}_{\pi{}\pi{}}^I$ corresponds to the isospin-$I$ channel.
These matrix elements will be evaluated in sect.~\ref{sec:njl}
within the NJL model. 

For a local Bose distribution
with temperature $T$ and pion chemical potential $\mu_\pi$,
\beq
\label{eq:short-nBx}
f^{(0)}_p = 
\frac{1}{
\exp \left[\left( p^\nu u_\nu(x) - \mu_\pi(x) \right) /T(x)\right] - 1}
\eeq
the collision term vanishes, {\it i.e.}, $C_{\pi \pi}[f^{(0)}]=0$.
Because of this, the system is then said to be in `local thermal 
equilibrium'. In general, however, $f^{(0)}_p$ is not a solution of the 
BUU equation, since the derivatives on the left-hand side of 
eq.~(\ref{eq:short-BUU}) do not vanish for $x$-dependent fields.

We are particularly interested in static solutions\footnote{
Since we want to calculate static transport coefficients, we should
consider static solutions. 
In principle, we can set up the system to be in local thermal equilibrium 
at some given time $t_0$. However, the BUU equation then drives the
system away from the equilibrium distribution, {\it i.e.}, this solution would not 
be static. 
}
with a non-vanishing shear flow.
To this end, we work in the local rest frame of the fluid,
$u = (1,0,0,0)^T$, and consider a time-independent equilibrium distribution  
$f^{(0)}$ with $\partial^i u^j \neq 0$ for some spatial indices $i \neq j$.
Moreover, we assume that $T$ and $\mu_\pi$ do not depend on $x$.
The solution of the the BUU equation is then constructed in the framework 
of a Chap\-man-Enskog expansion 
\cite{ChapmanCowling:MTNUG,dGvLvW:RKT,Liboff:KT}
to leading order. 
This means, we write
\beq
f_p = f^{(0)}_p +  f^{(1)}_p,
\label{eq:fpiexpansion}
\eeq
and linearize the collision term $C^{\pi\pi}$ in the correction $f^{(1)}_p$,
while the advective term (the left-hand side of eq.~(\ref{eq:short-BUU}))
is evaluated with $f^{(0)}_p$ only. Parametrizing $f^{(1)}_p$ as 
\beq
    f_p^{(1)} 
    = f_p^{(0)}(1+f_p^{(0)})\, B_p^{ij}\,\tau^{\mathit{shear}}_{ij}
\eeq
where
$\tau^{\mathit{shear}}_{ij} = 
\frac{1}{2}(\partial_iu_j + \partial_j u_i - \frac{2}{3}\delta_{ij} 
\vec\nabla\cdot \vec u)$
in the local rest frame of the fluid, one finally arrives at 
\begin{align}
&- f^{(0)}_p \left( 1+ f^{(0)}_p \right)
\frac{p^i p^j}{ET}
\;=
\nonumber \\ & 
\frac{g_\pi}{2}
  \intddd{p'} \intddd{p_1}  \intddd{p_1'}
  \;
  (2\pi)^4\delta(^4p+p_1-p'-p_1') \, \times
\nonumber \\ 
&  \hspace{16mm} \times
  \frac{|{\cal{M}}_{\pi \pi}|^2 }{16EE_1E'E_1'}\,
  f^{(0)}_p f^{(0)}_{p_1}
  (1+f^{(0)}_{p'})(1+ f^{(0)}_{p_1'}) \, \times 
\nonumber \\ 
& \hspace{16mm} \times
  \left[ B^{ij}_{p'} + B^{ij}_{p_1'} - B^{ij}_p - B^{ij}_{p_1} \right],
\label{eq:BUUlin}
\end{align}
which is a linear integral equation for the tensor function
\beq
B_p^{ij} =
{\cal{B}}(|\vecc{p}|)\left(\hat{p}^i\hat{p}^j
-\frac{1}{3}\delta^{ij}\right),
\label{eq:Bp}
\eeq
where $\hat{p} = \vec{p}/{|\vec{p}|}$.
We solve this equation by expanding the function ${\cal{B}}$ in
generalized Sonine polynomials. Further details are given
in the appendix.

Having constructed the distribution function $f$,
we can now insert it into eq.~(\ref{eq:Tmunu_fpi})
to determine the energy-moment\-um tensor. The shear viscosity
is then obtained by comparing the result with Eqs.~(\ref{eq:T_munu}) --
(\ref{eq:T_munu1}). In fact, as the equilibrium distribution 
just gives rise to the ideal part of $T^{\mu\nu}$, the relevant 
contribution comes from the correction term $f^{(1)}$. 
One finds
\beq
\label{eq:short-rkt-eta-B}
\eta =  \frac{g_\pi}{15}  \frac{4\pi}{(2\pi)^3}
\int\limits_0^{\infty} \intd p \,\frac{p^4}{E}
f_p^{(0)}(1+ f_p^{(0)})\, {\cal{B}}(p).
\eeq

Finally we note that
the kinetic description is only valid in the dilute-gas limit,
{\it i.e.}, if the system is well approximated by a free gas and
interactions lead only to minor corrections.
This means that the mean free path $\lambda$ has to be much larger than 
the typical range of interaction $\dorria$. To assess the region of
validity of our results we therefore have to check whether this condition is 
fulfilled.


\section{In-medium pion scattering in the NJL model}
\label{sec:njl}

The remaining task for the calculation of the shear viscosity in the
approximation, discussed above, is the determination of the in-medium 
pion mass $m_\pi$
and the $\pi{}\pi$-scattering amplitude ${\mathcal{M}}_{\pi{}\pi}$.
As motivated in the Introduction, this will be done within the NJL model,
which has the advantage that it allows for calculating medium-dependent
meson properties, including effects of their quark substructure and
the chiral transition.

In the present analysis we restrict ourselves to two quark flavors
and use the standard NJL-model Lagrangian \cite{NJL:I,NJL:II},
\beq
    {\cal L}
    =
    \overline{\psi}(i \slashed{\partial} -m_0)\psi
    +
    g[(\overline{\psi}\psi)^2+(\overline{\psi}i\gamma_5\vec{\tau}\psi)^2],
\eeq
with four-point interactions in the scalar-isoscalar ($\sigma$) and
pseudoscalar-isovector ($\pi$) channels. Here $\psi$ denotes a quark
field, $m_0$ is the bare quark mass, and $g$ is a coupling constant
with mass dimension $-2$. The model is not renormalizable and we
use a 3-momentum cut-off $\Lambda$ in order to regularize divergent
loop integrals.
For the numerical calculations we take the parameters $m_0=5.0$~MeV,
$g\Lambda^2=2.06$, and $\Lambda=664.3$~MeV from \cite{Buballa:Review05}.

\subsection{Pion mass}

\begin{figure}[tb]
  \begin{center}
    \begin{displaymath}
      \parbox{11ex}{\includegraphics[height=9ex]{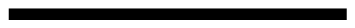}}
      \quad
      =
      \quad
      \parbox{11ex}{\includegraphics[height=9ex]{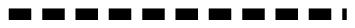}}
      \quad
      +
      \quad
      \parbox{16ex}{\includegraphics[height=9ex]{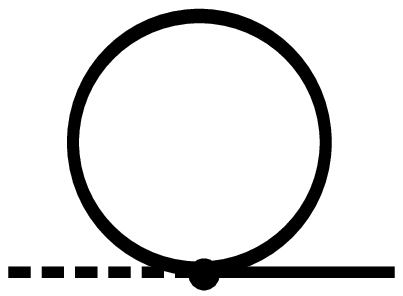}} .
      \end{displaymath}
      \caption{Gap equation in the Hartree approximation. The dashed
        line corresponds to the bare quark propagator $S_0$,
        the solid line corresponds to the dressed quark propagator $S$.}
    \label{fig:gapeq}
  \end{center}
\end{figure}

The strong attractive interaction between the quarks leads
to the dynamical generation of a mass gap.
In the Hartree approximation, this so-called constituent quark mass
is given by 
\beq
m_q = m_0 +\Sigma_H,
\eeq
where $\Sigma_H$ is the one-loop self-energy, diagrammatically
depicted in fig.~\ref{fig:gapeq}.
We refer to the literature for the exact expression of $\Sigma_H$, 
see e.g. \cite{Oertel:2000jp} for details.
Note that unlike the bare mass $m_0$, $\Sigma_H$ and, thus, $m_q$
are temperature dependent, so that typically $m_q \gg m_0$ at low
$T$, whereas $m_q \approx m_0$ at high $T$.

\begin{figure}[tb]
  \begin{center}
    \begin{displaymath}
    \parbox{14ex}{\includegraphics[height=9ex]{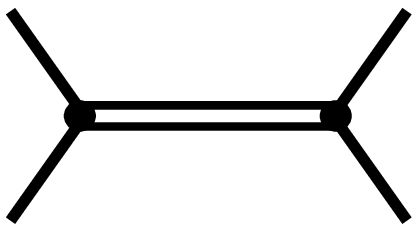}}
    =
    \quad
    \parbox{7ex}{\includegraphics[height=9ex]{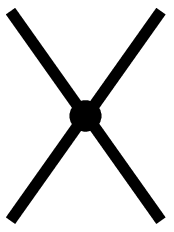}}
    +
    \quad
    \parbox{20ex}{\includegraphics[height=9ex]{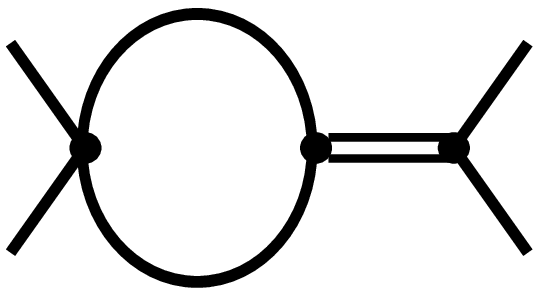}}
      \end{displaymath}
      \caption{Bethe-Salpeter equation for quark-antiquark scattering
      in Random-Phase Approximation.
      The $T$-matrix is depicted by double lines, while single lines
      denote the dressed quark propagator.}
    \label{fig:bse}
  \end{center}
\end{figure}

Quark-antiquark scattering processes in specific channels can be
associated with the propagation of a meson.
In the Random-Phase Approximation (RPA) the Bethe-Salpeter equation for the
quark-antiquark $T$-matrix takes the form, diagrammatically shown
in fig.~\ref{fig:bse}.
This approximation is a standard technique in the NJL model, 
again we refer to \cite{Oertel:2000jp} for details of the calculation.
Evaluation of this equation in the scalar and pseudoscalar channels,
respectively, yields the meson ``propagators''
\beq
\label{eq:mesonprop}
D^{RPA}_M(q) = \frac{-2g}{1-2g\Pi^{RPA}_M(q)}.
\eeq
where $M\in\{\pi^a, \sigma\}$,
and $\Pi_M$ denotes the corresponding quark-antiquark polarization loop.
For the momenta we have used the short-hand notation
$q \equiv (i\omega_m,\vecc{q})$ with bosonic
Matsubara frequencies $\omega_m = 2m\pi T$.
After analytic continuation to real energies, 
which simply amounts to replacing $i\omega_m$ by $q_0+i\varepsilon$ in the 
final expressions,\footnote{This corresponds to the retarded propagator.
Alternatively, we could replace 
$i\omega_m$ by $q_0+i\,\mathrm{sign}(q_0)\varepsilon$
for the time-ordered propagator.}
the meson masses $m_M$ can be
extracted from the position of the pole of $D_M$,
\beq
\label{eq:mesonmass}
1 - 2 g \mathrm{Re}\, \Pi^{RPA}_M(q_0=m_M, \vecc{0}) = 0.
\eeq
Here we have chosen to define the ``mass'' of an unbound resonance
via the real part of $\Pi_M$, while for bound-state solutions, {\it i.e.},
for $q_0 = m_M < 2m_q$ the polarization function is real anyway.

\begin{figure}
  \begin{center}
      \psfrag{m}{\hspace{-2ex}{{$m\,[\MeV]$}}}
      \psfrag{t}{\hspace{-3ex}{{$T [\MeV]$}}}
      \psfrag{Tdiss}{\hspace{-0ex}{\small{$T_{diss}$}}}
      \psfrag{Tmott}{\hspace{-0ex}{\small{$T_{Mott}$}}}
      \psfrag{q}{\hspace{-.5ex}{\small{$q$}}}
      \psfrag{pi}{\hspace{-0ex}{\small{$\pi$}}}
      \psfrag{si}{\hspace{-0ex}{\small{$\sigma$}}}
      \includegraphics[height=\columnwidth,
        angle=-90]{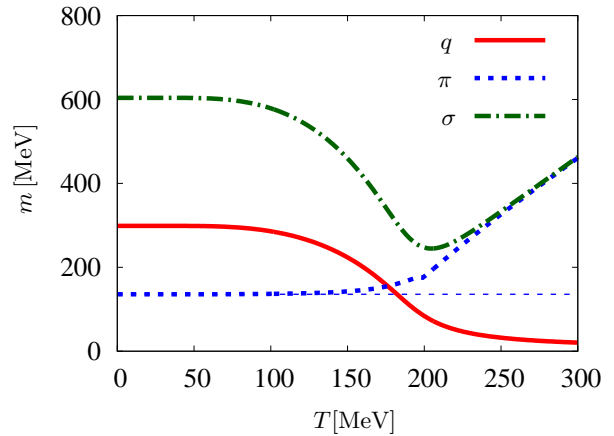}
      \caption{Masses of constituent quarks (solid/red line), pions
        (dotted/blue line), and
        $\sigma$-mesons (dashed-dotted/green line) as functions of temperature. The vacuum-pion
        mass is shown as a thin line.}
      \label{fig:masses}
  \end{center}
\end{figure}
Like the constituent quark mass, the masses of the pion and the
$\sigma$-meson are temperature dependent.
This is well known from the literature (see, {\it e.g.},
\cite{Klevansky:Review92}) and shown in fig.~\ref{fig:masses} for our
choice of parameters. The latter have been fitted to reproduce a pion
mass of $m_\pi=135~\MeV$ in vacuum~\cite{Buballa:Review05}.
In the chirally broken phase, the pion is protected by being an approximate
Goldstone boson, and hence the temperature dependence of $m_\pi$ is very
weak.
The mass of the $\sigma$-meson, on the other hand, is approximately given by
$2m_q$ and therefore drops considerably when approaching the crossover
temperature. In the restored phase, $m_\pi$ and $m_\sigma$ rise again and
become asymptotically degenerate.

For the later discussion of the $\pi{}\pi$-scattering cross section
it is useful to define two characteristic temperatures in the
crossover region, which have been introduced first in
ref.~\cite{QuackKlevansky-etal:PLB95}.
The first one is the
$\sigma$-dissociation temperature, defined by
\beq
m_\sigma (T_{diss}) = 2 m_\pi(T_{diss}).
\eeq
For $T<T_{diss}$, the $\sigma$-meson can decay into two pions.
The second is the Mott-temperature, defined by
\beq
m_\pi (T_{Mott}) = 2 m_q(T_{Mott}).
\eeq
For $T>T_{Mott}$, the pion can decay into a quark-antiquark pair, {\it i.e.},
it is no longer a bound state but only a $q \bar{q}$ resonance.
With the parameter set chosen here, we have $T_{diss}=179.9\, \MeV$ and
$T_{Mott}=198.55\,\MeV$. In the chiral limit, both temperatures are equal to 
the critical temperature of the chiral phase transition. Thus, they can be taken as
possible definitions of the chiral crossover temperature.

\subsection{$\pi{}\pi$-scattering length}
\label{sec:a-pipi}

\begin{figure}
  \begin{center}
    {\includegraphics[height=12ex]{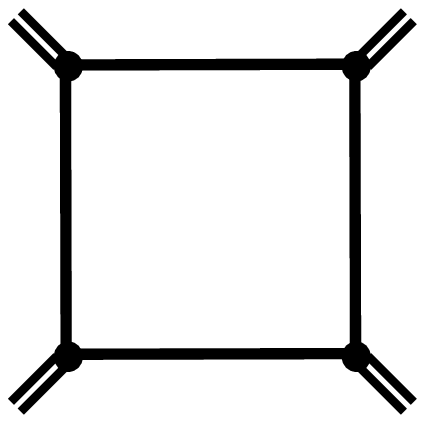}}
    \hspace{8ex}
    {\includegraphics[height=12ex]{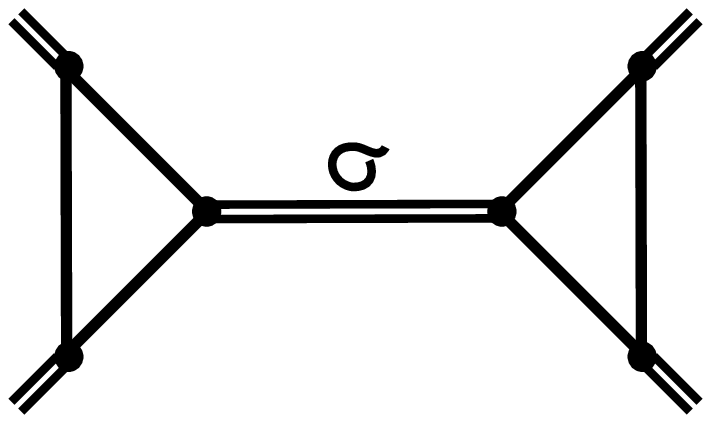}}
  \end{center}
  \caption{Diagrams contributing to $\pi{}\pi$-scattering in leading-order 
$1/N_c$. Left: box diagram. Right: $\sigma$-meson exchange.
In the quark loops, both directions of the momentum flow should be taken
into account.}
  \label{fig:pipi-scattering-diagrams}
\end{figure}

Since pions are not elementary degrees of freedom in the NJL model, there is
no elementary pion-pion interaction in the model. However, effective
meson-meson vertices can be constructed by using the
diagrammatic building-blocks of the NJL model. We follow
\cite{Schulze:JPG95,QuackKlevansky-etal:PLB95} and consider the
leading-order diagrams in a $1/N_c$-expansion. The two diagrams
contributing to ${\cal{M}}_{\pi \pi}$ to that order are shown in
fig.~\ref{fig:pipi-scattering-diagrams}.
In the first diagram the pions are coupled through an intermediate quark
loop (`box'), while the second corresponds to the exchange of a
$\sigma$-meson, which is coupled to the external pions via quark
triangles.\footnote{
In the limit of infinitely heavy quarks,
the quark box and the triangle shrink to a pointlike four-pion
and $\sigma\pi\pi$-vertex, respectively. The diagrams shown in
fig.~\ref{fig:pipi-scattering-diagrams} then reduce to the leading-order
contributions for $\pi{}\pi$-scattering in the linear sigma
model~\cite{Gasiorowicz:1969kn}.}
To be precise, the diagrams shown in fig.~\ref{fig:pipi-scattering-diagrams}
are representatives of two classes of diagrams which must be
considered~\cite{Schulze:JPG95}.
In the case of the box diagram, the pions can be attached to the quark
loop in various orderings, leading to three distinct amplitudes.
Moreover, the $\sigma$-meson exchange must be taken in the $s$-, $t$- and
$u$-channel. These contributions can be coupled to the different isospin
channels.
The ${\pi{}\pi}$-scattering length is
related to the invariant matrix element at threshold,\footnote{
Since we do not expect any confusion,
we denote both, the center-of momentum energy and the entropy density, 
by $s$, as customary.
} 
\beq
\label{eq:aI}
a^I = \frac{1}{32 \pi m_\pi}{\cal{M}}_{\pi \pi}^I(s=4m_\pi^2, t=u=0).
\eeq
The explicit expressions for $a^0$ and $a^2$ as functions of
temperature are given in ref.~\cite{QuackKlevansky-etal:PLB95},
while $a^1$ vanishes, as it should for $s$-wave scattering.

\begin{figure}
  \begin{center}
      \psfrag{a}{\hspace{-3ex}{\small{$a\,[m_\pi^{-1}]$}}}
      \psfrag{t}{\hspace{-2.5ex}{\small{$T [\MeV]$}}}
      \psfrag{Tdiss}{\hspace{-0ex}{\small{$T_{diss}$}}}
      \psfrag{TMott}{\hspace{-0ex}{\small{$T_{Mott}$}}}
      \psfrag{a0}{\hspace{-.5ex}{\small{$a^0$}}}
      \psfrag{a2}{\hspace{-.5ex}{\small{$a^2$}}}
        \includegraphics[height=\columnwidth, angle=-90]{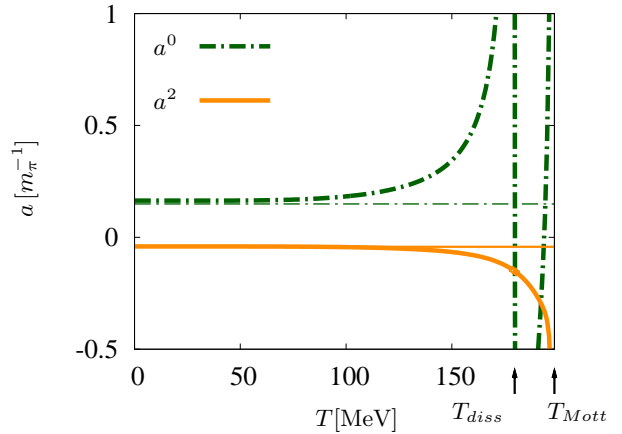}
  \end{center}
  \caption{The $\pi{}\pi$-scattering lengths as functions of
    temperature. The dash-dotted (green) lines correspond to $I=0$,
    whereas the solid (orange) lines correspond to $I=2$.
    Bold lines denote the NJL results, whereas the thin lines
    correspond to the vacuum-extrapolated Weinberg scattering lengths.}
  \label{fig:scattering-length}
\end{figure}

The results are shown in fig. \ref{fig:scattering-length}.
With the parameter set used in this article, we find the same qualitative
behavior as in \cite{QuackKlevansky-etal:PLB95}.
The vacuum values
$a^0 = 0.164\, m_\pi^{-1}$ and $a^2 = -0.041\, m_\pi^{-1}$,
are in reasonable agreement with the Weinberg scattering
lengths~\cite{Weinberg:PRL66}, 
\beq
\label{eq:M-Weinberg}
a_W^{0}=\frac{7 m_\pi}{32\pi f_\pi^2}
, \quad
a_W^{2}=-\frac{2 m_\pi}{32\pi f_\pi^2}.
\eeq
which yield $a^0_W = 0.149\, m_\pi^{-1}$ and $a^2_W = -0.0425 \, m_\pi^{-1}$.
Note that the chiral expansion employed by Weinberg is not identical to
the $1/N_c$ expansion. Thus perfect agreement should not be expected.

At finite temperature, the scattering lengths deviate from their vacuum
values. This effect is very small at low temperatures but becomes 
important when approaching the crossover region, in particular the two
transition temperatures $T_{diss}$ and $T_{Mott}$.
At $T = T_{diss}$, the process $\pi + \pi \to \sigma$ is resonant for two pions
at rest. Thus the $s$-channel $\sigma$-exchange diagram involves an
on-shell $\sigma$-propagator, giving rise to a strong peak in the isoscalar
channel.\footnote{Note that $a^0$ does not diverge at $T = T_{diss}$
since in the NJL model the $\sigma$-meson has a small but nonvanishing width
due to $q\bar q$-decay.
}
Similarly, at $T=T_{Mott}$, a pion
at rest is at threshold for the decay process into a quark-antiquark pair.
This leads to a divergent scattering length in both isospin channels at
the Mott temperature

The resonant scattering length at $T_{diss}$ and $T_{Mott}$
is reminiscent of the behavior of cold atomic gases in an external magnetic
field in the vicinity of a Feshbach resonance.
In the present case the role of the magnetic field is taken by
the temperature.
This analogy between cold atoms and hot pions will be investigated further
in the following sections.

\subsection{Scattering amplitude}
\label{sec:Mpipi}

As discussed in sect.~\ref{sec:short-rkt},
the calculation of the shear viscosity
of an interacting pion gas within kinetic theory involves
the in-medium scattering amplitudes
${\cal{M}}_{\pi{}\pi{}}^I$.
In the following we introduce four approximate ways to evaluate
${\cal{M}}_{\pi{}\pi{}}^I$. 
Comparison of the corresponding results will then allow to identify 
the most relevant effects.

\subsubsection{Weinberg amplitude}
\label{sec:Mpipi-Weinberg}

The most simple approximation is to neglect both, momentum and
temperature dependence of the scattering amplitude, and to employ
Eqs.~(\ref{eq:aI}) and (\ref{eq:M-Weinberg})
to construct the latter from the Weinberg scattering 
lengths,
\beq
{\cal{M}}_{\pi{}\pi{},W}^{0} = \frac{7m_\pi^2}{f_\pi^2}
, \quad
{\cal{M}}_{\pi{}\pi{},W}^2 = -\frac{2m_\pi^2}{f_\pi^2}
\eeq
This approximation has previously been discussed in 
ref.~\cite{DLlE:04}.

\subsubsection{Medium modified momentum independent amplitude}

We have seen in sect.~\ref{sec:a-pipi}
that the assumption of a temperature independent scattering amplitude is
definitely not a good approximation when approaching the crossover region.
Therefore, in order to improve on this, we replace the Weinberg scattering 
lengths by the temperature dependent scattering lengths of
fig.~\ref{fig:scattering-length}.
However, we still neglect the momentum dependence,
{\it i.e.}, we evaluate the amplitudes only at the two-pion threshold, 
\beq
    {\cal{M}}_{\pi{}\pi{},th}^{I}
    =
    {\cal{M}}_{\pi \pi}^I(s=4m_\pi^2, t=u=0).
\eeq
At low temperature the Weinberg amplitudes are approximately
recovered because of the good agreement of the NJL scattering lengths with
the Weinberg ones.
On the other hand, ${\cal{M}}_{\pi{}\pi{},th}^{I}$ also incorporates the
``Feshbach resonances'' at $T_{diss}$ and $T_{Mott}$. Of course, this will
be important for the shear viscosity.

\subsubsection{Momentum dependent $\sigma$-meson exchange}

On the other hand, in the crossover region where the temperature is of the
order of the pion mass, the approximation of a generally momentum dependent
scattering amplitude by its value at the two-pion threshold becomes
questionable as well.
This is most obvious for the $\sigma$-meson exchange diagram 
(Fig.~\ref{fig:pipi-scattering-diagrams}, right)
in the $s$-channel.
As explained above, the resonant behavior of $a^0$ at $T=T_{diss}$ is due
to the fact that at this temperature the $\sigma$-meson propagator is
on-shell for pions at rest. However, when the thermal motion of the pions
is taken into
account, the $\sigma$-meson can become on-shell already at lower
temperatures, while at $T=T_{diss}$ a large fraction of pion pairs is
far away from the threshold.

In order to include this effect, we again improve
our approximation scheme and take the momentum dependence of the
$\sigma$-propagator into account. To be precise, in the $s$-channel
we consider a $\sigma$-meson with energy $q^0 = \sqrt{s}$ and vanishing
three-momentum, while in the $t$- and $u$- channel we take
$q^0 = 0$ and $|\vecc q| = t$ or $u$, respectively.
For simplicity, we still neglect the momentum dependence of the
quark triangle and box diagrams.
This approximation is sometimes called  the `static limit' and
can be interpreted as not resolving the quark substructure of the effective
meson-meson vertices. We expect that this is a minor effect, except for 
temperatures close to and above $T_{Mott}$. 

For small momenta, the resulting scattering amplitude is equal to
${\cal{M}}_{\pi{}\pi{},th}^{I}$, so the Weinberg
scattering length is again well reproduced at $T=0$.
Above the two-pion threshold there is now also a small $p$-wave contribution,
giving rise to a non-vanishing amplitude in the isospin-1 channel.
This is taken into account in our calculations. However, for a realistic
description of this channel, the $\rho$-meson has to be included.
This extention of the model is left for future work.

\subsubsection{Including the $\pi{}\pi$-decay width of the  $\sigma$-meson}
\label{sec:Mpipi-dressed}

\begin{figure}
  \begin{center}
    {\includegraphics[height=12ex]{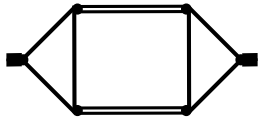}}
  \end{center}
  \caption{Selfenergy correction to the $\sigma$-meson propagator
   with two intermediate pions.}
  \label{fig:sigma-pipi-loop}
\end{figure}

A still unrealistic feature of the previous approximation is the
structure of the intermediate $\sigma$-meson.
While the process $\pi\pi\leftrightarrow\sigma$ is an essential
ingredient of the $\pi{}\pi$-scattering amplitude, the coupling of the
$\sigma$-meson to a two-pion intermediate state is not included
in the RPA Bethe-Salpeter equation, depicted in fig.~\ref{fig:bse}.
As a consequence, the $\sigma$-meson only has a small width due to
quark-antiquark decay, whereas the $\pi\pi$-decay width, which should
be dominant at low temperature, is not taken into account.

Thus, in an improved approximation we consider a dressed
$\sigma$-propagator,
\beq
D^{dressed}_{\sigma}(q) =
\frac{-2g}{
1 - 2 g \Pi^{dressed}_\sigma(q)},
\label{eq:Ddressed}
\eeq
where $\Pi^{dressed}_\sigma(q)$ is a sum of the RPA polarization loop
and a correction term which contains a two-pion intermediate state.
This correction can be derived systematically in an expansion in the
inverse number of colors ($1/N_c$) to next-to-leading order~\cite{Oertel:2000jp,Quack:1993ie,Dmitrasinovic:1995cb,Oertel:1999fk,Oertel:2000sr,Plant:2000ty,Goeke:2007bj}.
The relevant diagram is shown in fig.~\ref{fig:sigma-pipi-loop}
and yields
\beq
\Pi^{\pi\pi}_\sigma (q) =
- \frac{3}{2} \Gamma_{\sigma \pi \pi}^2
T \sum_{n} \int \! \frac{d^3k}{(2\pi)^3}
D_{\pi}^{RPA} (q+k) D_{\pi}^{RPA} (k),
\eeq
with $q = (i\omega_m, \vecc{q})$, $k = (i\omega_n, \vecc{k})$
and bosonic Matsubara frequencies $\omega_m$ and $\omega_n$.
$\Gamma_{\sigma \pi \pi}$ denotes the effective $\sigma\pi\pi$
vertices, {\it i.e.}, the quark triangles in fig.~\ref{fig:sigma-pipi-loop},
which are again evaluated in the static limit.

After analytical continuation to real energies, $\Pi^{\pi\pi}_\sigma$
has a nonvanishing imaginary part above the two-pion threshold.
This is exactly the wanted effect.
In addition, it has also a real part, leading to a renormalization
of the $\sigma$-meson mass.
Unfortunately, this causes inconsistencies with chiral low-energy theorems,
which are based on a delicate cancellation between the $\sigma$-exchange
and the box diagram in fig.~\ref{fig:pipi-scattering-diagrams}.

This problem cannot easily be cured. 
In principle, consistency with chiral symmetry can be achieved in a 
systematic $1/N_c$ expansion by taking into account all diagrams of
a given order. At leading order these are the diagrams shown 
in fig.~\ref{fig:pipi-scattering-diagrams} with the RPA $\sigma$-meson, 
while at next-to-leading order there are more than 60 additional diagrams,
not counting the different orderings to attach the external pions. 
Moreover, the fact that fig.~\ref{fig:sigma-pipi-loop} is 
iterated in the $\sigma$-propagator
spoils the strict $1/N_c$-counting, and we would encounter inconsistencies
even if all next-to-leading order diagrams were considered. 

In this situation, we decided to include the $\sigma$-width in a minimal
fashion by considering only the imaginary part of $\Pi^{\pi\pi}_\sigma$
but neglecting the real part, {\it i.e.}, we take
\beq
\Pi^{dressed}(q) =
\Pi^{RPA}_\sigma(q) + i \mathrm{Im}\, \Pi^{\pi\pi}_\sigma (q).
\eeq
As a consequence, the propagator is not modified at and below the
two-pion threshold, which implies that the scattering lengths remain 
unchanged. In particular, the Weinberg scattering lengths
are still reproduced well at $T=0$.

Of course, this way of including the $\sigma$-width is very schematic.
At the present stage we mainly want to estimate its importance for the
shear viscosity, while we are not aiming a perfect description of 
$\pi\pi$-scattering.
We note that neglecting the real part of $\Pi^{\pi\pi}_\sigma$, 
which is related to the imaginary part by the Kramers-Kronig relation,
violates causality.
We believe, however, that this is a minor effect in comparison with 
the violation of the chiral low-energy theorems, which would occur
if the real part was taken into account. 

\begin{figure}[h!]
  \psfrag{rhosi}{$\rho_\sigma [\MeV^{-2}]$}
  \psfrag{q0}{\hspace{-2ex}$q_0\, [\MeV]$}
  \psfrag{T0}{\hspace{-2ex}$T=0\, \MeV$}
  \psfrag{T150}{\hspace{-2ex}$T=150\, \MeV$}
  \psfrag{T177}{\hspace{-2ex}$T=177\, \MeV$}
  \psfrag{T188}{\hspace{-2ex}$T=188\, \MeV$}
  \includegraphics[height=\columnwidth, angle=-90]{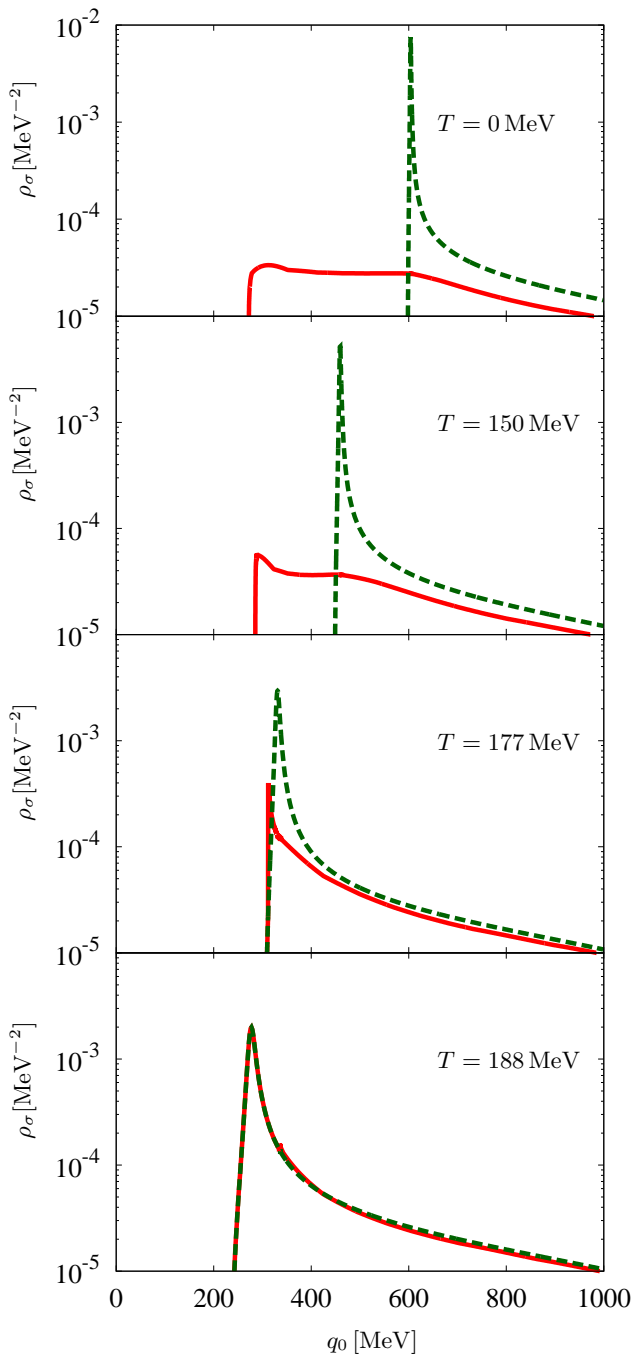}
  \vspace{8ex}
  \caption{Unnormalized spectral functions of the $\sigma$-meson as functions of energy $q_0$ for different temperatures. The dashed (green) lines indicate the RPA calculations, while the solid (red) lines include dressing of the $\sigma$-meson as described in the text.}
  \label{fig:rhosi}
\end{figure}

In fig.~\ref{fig:rhosi}, the spectrum of the $\sigma$-meson is shown for
different temperatures.
The ``spectral function''\footnote{The meson propagators
as defined in Eqs.~(\ref{eq:mesonprop}) and (\ref{eq:Ddressed}),
and, thus, the spectral functions defined in eq.~(\ref{eq:specfun})
are not properly normalized.
However, since it is the unrenormalized $\sigma$-propagator which enters 
into the $\pi\pi$-scattering matrix elements, we prefer to
compare the corresponding unnormalized spectral functions in 
fig.~\ref{fig:rhosi}.}
\beq
\label{eq:specfun}
\rho_\sigma(q) = -2\mathrm{Im}\, D_\sigma(q)
\eeq
is shown for $|\vecc{q}|=0$ as a function of energy for both approximations,
RPA as well as the additional dressing. In the vacuum, the RPA meson has a very
small spectral width, while the dressed $\sigma$-meson is much broader.
When the temperature increases, the quark mass decreases and thus
the threshold for the RPA meson moves downwards. At the same time, the mass
peak of the $\sigma$ moves downwards as well (see fig.~\ref{fig:masses}).
Therefore the phase space for the two-pion decay becomes smaller and the
width of the dressed $\sigma$-meson gets reduced.
At $T=177$~MeV, which is near the dissociation temperature,
the thresholds $2m_\pi$ and $2 m_q$ are almost equal.
At temperatures above the $T_{diss}$, the pion decay plays only a minor
role, and the dressed $\sigma$-meson is very similar to the RPA result.
The threshold in this temperature range is determined by the $q\bar q$-decay
channel, which is included in both descriptions of the $\sigma$-meson.

\subsection{Scattering cross section}

\begin{figure}
  \psfrag{sigtot}{$\sigma_{tot}\, [\millibarn]$}
  \psfrag{sqrts}{$\sqrt{s}\, [\MeV]$}
  \psfrag{T0}{$T=0\, \MeV$}
  \psfrag{T150}{$T=150\, \MeV$}
  \psfrag{T177}{$T=177\, \MeV$}
  \psfrag{T188}{$T=188\, \MeV$}
  \includegraphics[height=\columnwidth, angle=-90]{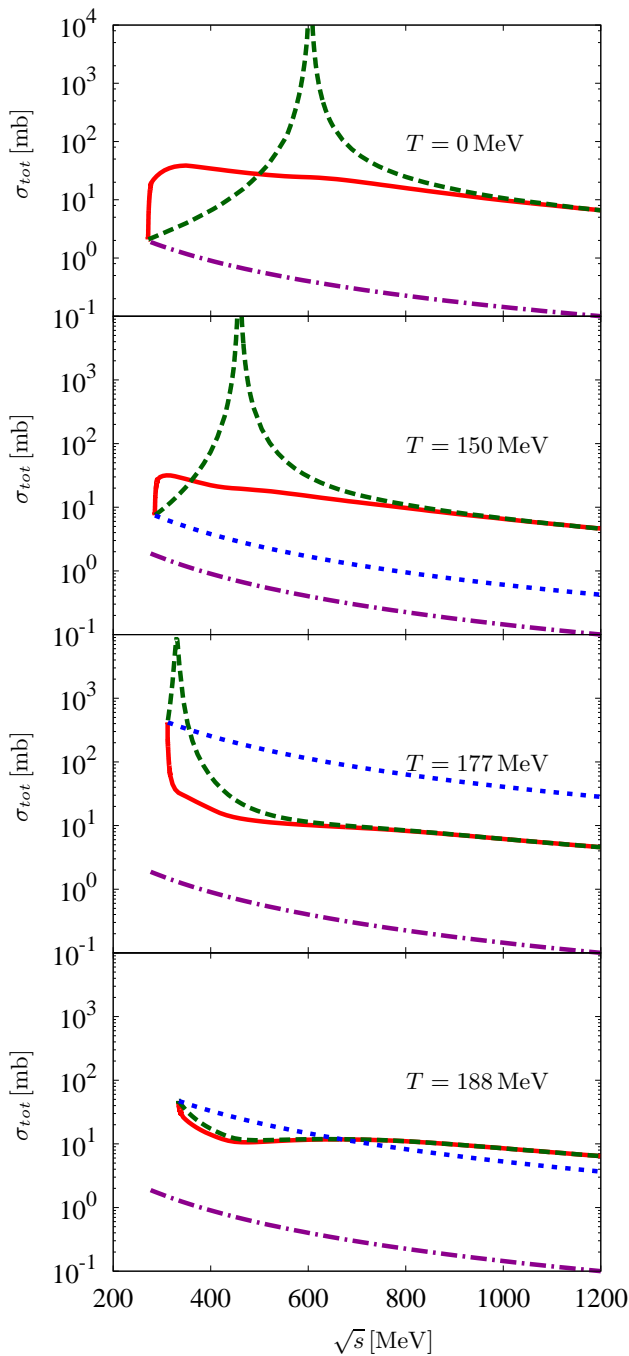}
  \vspace{8ex}
  \caption{Total cross section as a function of center-of-momentum energy
    $\sqrt{s}$ for four different temperatures. The different curves
    correspond to different approximations. The dash-dotted (purple) lines 
    were computed with the constant Weinberg scattering amplitudes 
    ${\cal{M}}_{\pi{}\pi{},W}^{I}$. The dotted (blue) lines correspond to the 
    temperature-dependent but momentum-independent amplitudes
    ${\cal{M}}_{\pi{}\pi{},th}^{I}$.
    Note that this line coincides with the previous one in vacuum. 
    For the results indicated by the dashed (green) lines, the momentum 
    dependence of the $\sigma$-meson propagator was taken into account
    in RPA, while the solid (red) lines were computed with the dressed $\sigma$.}
  \label{fig:sigtot-tdiff}
\end{figure}

The isospin-averaged differential cross section in the cen\-ter-of-mo\-men\-tum 
frame is given by
\beq
\left( \frac{\intd \sigma}{\intd \Omega} \right)_{cm}=
\frac{\left| {\cal{M}}_{\pi{}\pi{}} \right|^2}
{64 \pi^2 s},
\eeq
where $\left| {\cal{M}}_{\pi{}\pi{}} \right|^2$ is the isospin-averaged
squared scattering amplitude as defined in eq.~(\ref{eq:Mpipi}).
In fig.~\ref{fig:sigtot-tdiff} we compare the total cross sections of the 
four different approximations introduced in sect.~\ref{sec:Mpipi}
at four different temperatures: vacuum ($T=0$), 
an intermediate temperature ($T=150$~MeV),
$T=177$~MeV, which is closely below the $\sigma$-meson dissociation 
temperature, and $T=188$~MeV, which is between $T=T_{diss}$ and $T=T_{Mott}$.

For all temperatures, the three curves based on the NJL model coincide 
for $\sqrt{s}\to 2m_\pi$, but are in general quite different above threshold. 
In vacuum, since the NJL scattering lengths are in good 
agreement with the Weinberg values, the cross sections obtained with
the momentum independent amplitudes ${\cal{M}}_{\pi{}\pi{},th}^{I}$ 
and ${\cal{M}}_{\pi{}\pi{},W}^{I}$ (purple dash-dotted line) are almost 
identical. 
Taking into account the momentum dependence of the $\sigma$-meson propagator 
in RPA (green dashed line) leads to a very pronounced peak at 
$m_\sigma \approx 600\, \MeV$, 
which gets strongly broadened when the $\pi \pi$-decay width is included
(red solid line).

While the Weinberg cross section is temperature independent by 
construction, the NJL results change when the temperature is altered.
The main effects can be attributed to the temperature dependence of the 
$\sigma$-meson mass and, closely related to this, 
to the temperature dependence of the scattering lengths,
cf.~Figs.~\ref{fig:masses} and \ref{fig:scattering-length}.
When the scattering amplitude is approximated by its threshold value
(blue dotted lines), the latter simply leads to a scaling
of the vacuum result by an energy independent factor, which basically 
follows the behavior of the squared scattering lengths.  
In particular the whole curve (almost) diverges at the 
``Feshbach resonances'' $T=T_{diss}$ and $T=T_{Mott}$. 

When the momentum dependence of the $\sigma$-meson propagator is taken into 
account, the temperature effects are more subtle. 
Since the $\sigma$-meson mass decreases, the corresponding peak in the 
RPA calculation moves downwards in energy (green dashed lines), 
reaches the threshold at $T=T_{diss}$ and finally moves out of the  
kinematically allowed regime. 
Thus, whereas at $T=150$~MeV the qualitative behavior is still similar 
to the vacuum case, at temperatures close to or above the $\sigma$-meson 
dissociation temperature, the cross sections can fall considerably
below the approximation with the momentum independent amplitudes.
It is also remarkable that, except for the threshold region, the cross
sections obtained with the dressed $\sigma$-meson (red solid lines)
are rather temperature independent. 

Finally, we note that the NJL-model cross sections are typically orders of
magnitude larger than the cross sections obtained from the Weinberg
scattering lengths and agree with the latter only at low temperatures at 
threshold. Of course, this also has important consequences for the shear 
viscosity, which we discuss in the next section.

\section{Shear viscosity and the ideal fluid}
\label{sec:results}

The masses and scattering amplitudes computed in the previous section 
can now be used to calculate the shear viscosity as outlined in 
sect.~\ref{sec:short-rkt}, {\it i.e.}, from Eq.~(\ref{eq:short-rkt-eta-B}). 
Thereby we restrict ourselves to the case of
a vanishing pion chemical potential, $\mu_\pi = 0$.

\subsection{Shear viscosity}

 \begin{figure}
  \begin{center}
    \psfrag{eta}{\hspace{-3ex}{{$\eta\,[\GeV^3]$}}}
    \psfrag{t}{{{\hspace{-3ex}$T [\MeV]$}}}
    \psfrag{mfp}{\hspace{-1ex}{\small{mfp}}}
    \psfrag{Weinberg}{\hspace{-2.5ex}{\small{Weinberg}}}
    \includegraphics[height=\columnwidth,
      angle=-90]{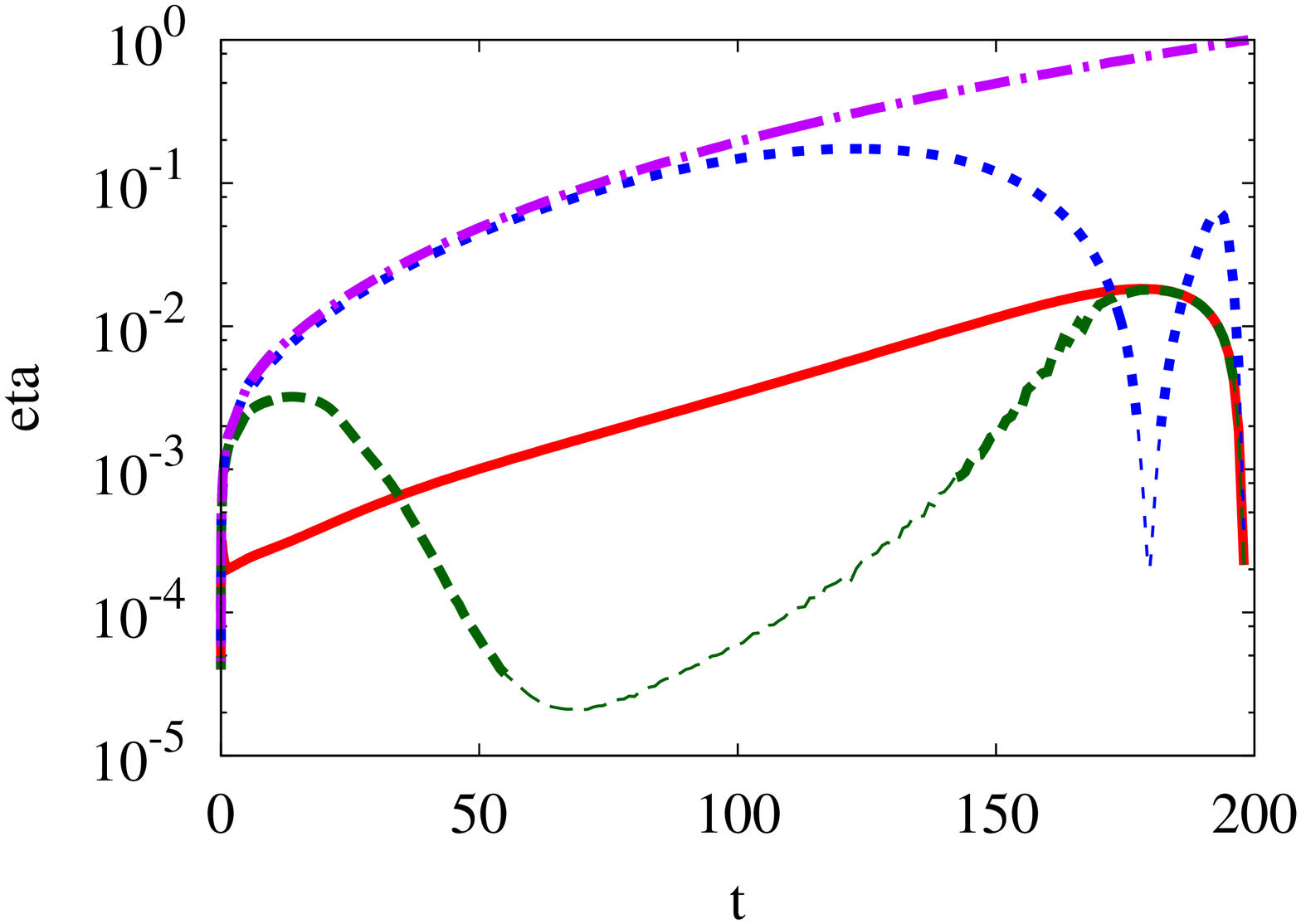}
    \includegraphics[height=\columnwidth,
      angle=-90]{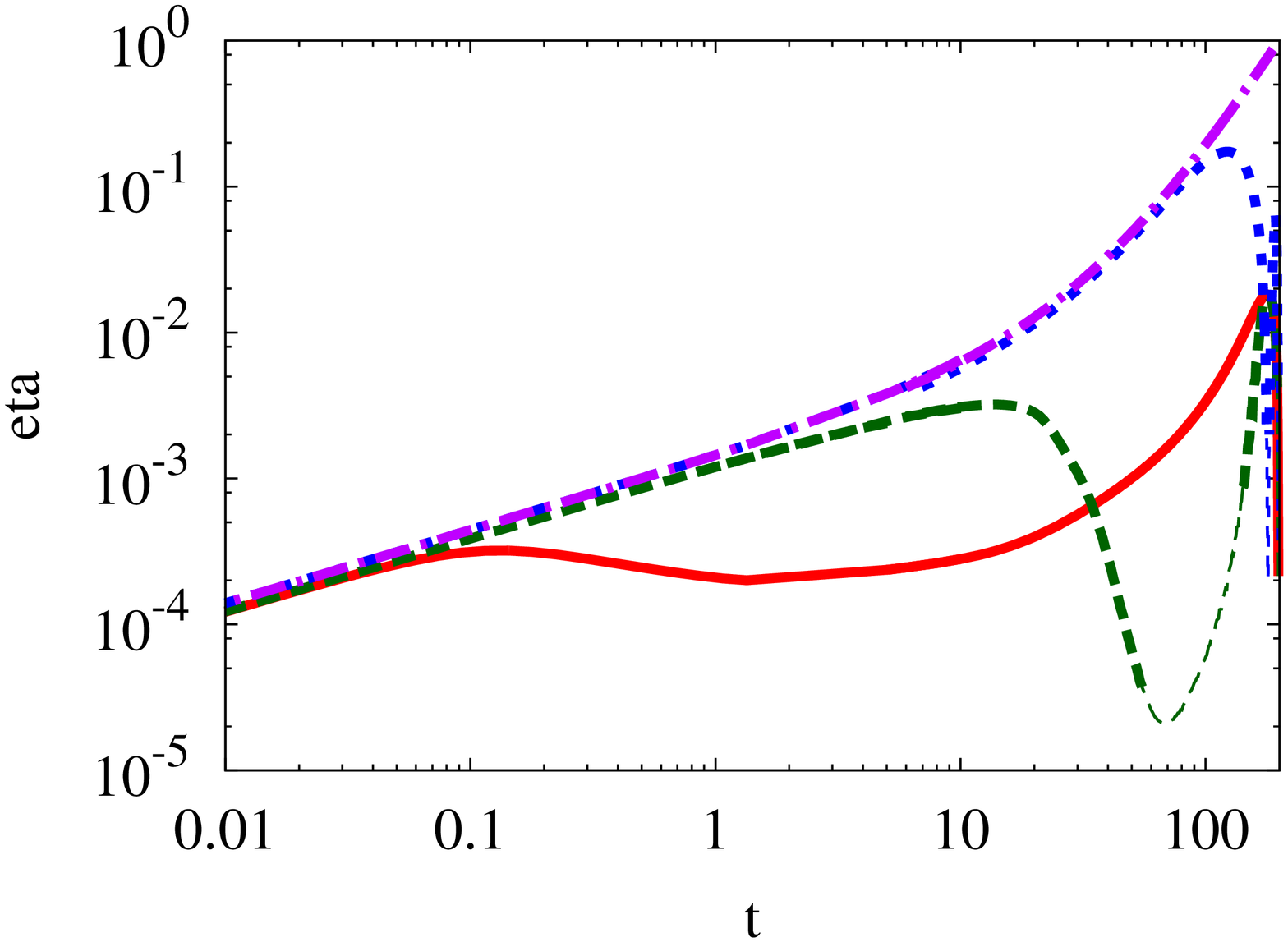}
    \caption{The shear viscosity $\eta$ as a function of temperature for the 
different approximations for the scattering amplitudes discussed in the text. 
The lower part of the figure shows the same results as the upper part, but 
in a log-log representation in order to highlight the low-temperature behavior.
The line styles are consistent with the choice in fig.~\ref{fig:sigtot-tdiff}: 
The dash-dotted (purple) line was obtained with the constant Weinberg 
scattering amplitudes ${\cal{M}}_{\pi{}\pi{},W}^{I}$ , the dotted (blue) line 
with the temperature dependent but momentum independent amplitudes
${\cal{M}}_{\pi{}\pi{},th}^{I}$. For the dashed (green) line, the momentum 
dependence of the $\sigma$-meson propagator was taken into account in RPA, 
while the solid (red) line was computed with the dressed $\sigma$.
Thin lines indicate the regions where the validity of the Boltzmann approach 
is questionable. Here we have defined these regions as the intervals where the 
lower boundary of the corresponding band in fig.~\ref{fig:validity} is less 
than unity.}
    \label{fig:eta}
  \end{center}
\end{figure}

The shear viscosity $\eta$ as a function of temperature $T$ is shown
in fig. \ref{fig:eta},
where again the results of the different approximations for the scattering
amplitude are compared.
For a qualitative interpretation it is useful to recall the simple
nonrelativistic approximation 
\beq
\eta \approx \bar p / (3\bar \sigma_{tot})~,
\label{eq:etaest}
\eeq
where $\bar p$ and $\bar \sigma_{tot}$ 
are the thermally averaged pion momentum and cross section, respectively. 
Since with increasing temperature, increasing momenta and energies are probed,
the cross section obtained with the constant Weinberg amplitude decreases
with temperature (cf. fig.~\ref{fig:sigtot-tdiff}) and therefore the 
shear viscosity rises monotonously in this approximation (purple dash-dotted
line).

Almost the same result is obtained with the momentum independent NJL amplitude
(blue dotted line) in the regime below $T\approx 100$~MeV, where this 
amplitude is in good agreement with the Weinberg value. 
On the other hand, in the vicinity of the chiral crossover, in particular
at $T_{diss}$ and $T_{Mott}$, the shear viscosity becomes very small because 
of the large cross section in this region.

When the momentum dependence of the RPA $\sigma$-meson propagator is taken 
into account (green dashed line), the minimum of the viscosity moves to
lower temperature.
In fact, we should expect that the minimum is roughly at the temperature
where the mean center-of-momentum energy matches the $\sigma$-meson mass peak.
This estimate yields $T_{min} \approx 100$~MeV, while the exact value
is even somewhat lower, $T_{min} \approx 70$~MeV.
Accordingly, at low and intermediate temperatures, the shear viscosity is 
considerably lower than in the previous approximations.  
On the other hand, when the dissociation temperature is approached, 
the energy of most pion pairs is way above the $\sigma$-meson mass peak and, 
hence, the viscosity is much larger than in the case of the momentum 
independent threshold amplitude.
Note, however, that the strong decrease towards the Mott temperature 
persists. This is most likely an artifact of neglecting the momentum
dependence of the quark triangles and box diagrams.

When we consider the dressed $\sigma$-meson, we obtain the result
indicated by the red solid line.
At low temperature the broadening of the mass peak leads to a strong 
enhancement of the cross section in the relevant kinematical region, whereas at 
intermediate temperatures, it is strongly reduced. 
As a consequence, the dip which is found in RPA, leading to a minimum 
at $T \approx 70$~MeV, is completely washed out.
Above the dissociation temperature, the pion decay channel of the 
$\sigma$-meson is of course irrelevant, and the shear viscosity practically 
coincides with the RPA result.

Finally, it should be noted that in the low-temperature limit all 
approximations approach the Weinberg result. This follows from the fact 
that at low enough temperature the momenta of the pions can be neglected 
and therefore only the scattering lengths are relevant. 
Hence the viscosities obtained from the momentum dependent scattering 
amplitudes converge to the result from the momentum independent
amplitude, which in turn is in good agreement with the Weinberg result.
It turns out, however, that in particular for the dressed $\sigma$-meson
one has to go to very low temperatures to see this behavior
(see lower part of fig. \ref{fig:eta}). Here we find that, because of 
the steep rise of the cross section near the threshold 
(see  fig.~\ref{fig:sigtot-tdiff}), the viscosity 
starts to deviate from the other approximations already at $T\approx 0.1$~MeV. 
Above this value, it first decreases and reaches a minimum at $T\approx 2$~MeV,
while the other curves strongly grow in this interval.  
As a consequence, the viscosities differ already by about an order of 
magnitude at $T = 5$~MeV.

Of course, these low temperatures are only of conceptual interest.
Since there is no pion chemical potential, the particle distribution
functions decrease exponentially with decreasing $T$ and become
extremely small.  
As a consequence we encountered numerical instabilities at temperatures 
below $5$~MeV. In this temperature regime, we have therefore performed
a nonrelativistic approximation in order to obtain numerically stable 
results.

\subsection{Validity of the kinetic description}

Although kinetic theory formally provides a viscosity 
for all interactions at any temperature, we should keep in mind that
the underlying assumption is the dilute gas limit, 
{\it i.e.}, the results can only be trusted if the mean free path $\lambda$ 
is much larger than the typical range of the interaction $\dorria$.
The mean free path is given by $\lambda=1/(n \bar\sigma_{tot})$,
where $n$ is the pion density, which can be calculated in the ideal-gas
limit. For the thermally averaged cross section we simply invert 
eq.~(\ref{eq:etaest}) and express it through the calculated shear 
viscosity. This yields $\lambda=3\eta/(n \bar{p})$.

In addition, we have to estimate the interaction range in an appropriate
way. Since in our model the long-range part of the $\pi\pi$-scattering is 
mediated by $\sigma$-meson exchange, we may set $\dorria =  1/m_{\sigma}$.
Two alternative estimates have been established in ref.~\cite{IOM:08}. 
The first is based on the hard-sphere limit, leading to
$\dorria = \sqrt{\bar\sigma_{tot}/\pi}$.
The second is the Compton wave length of the scattered
particle, {\it i.e.}, in our case $\dorria = 1/m_\pi$.

\begin{figure}
  \begin{center}
    \psfrag{lor}{\hspace{-0ex}{{$\lambda/r$}}}
    \psfrag{t}{{{\hspace{-3ex}$T [\MeV]$}}}
    \includegraphics[height=\columnwidth,
      angle=-90]{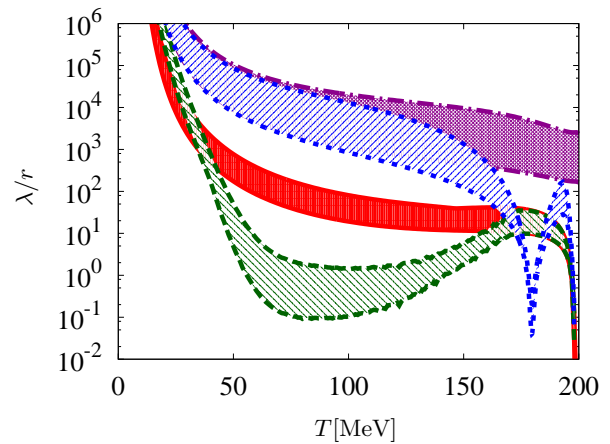}
    \caption{The ratio $\lambda/r$ as a function of temperature 
for the different approximations for the scattering amplitudes. 
The bands indicate the region between the results obtained with the largest 
and the smallest estimate for $r$. The colors and the line styles 
bordering the shaded regions are consistent with the choice in 
Figs.~\ref{fig:sigtot-tdiff} and \ref{fig:eta}. 
}
    \label{fig:validity}
  \end{center}
\end{figure}

Hence, since the definition of the interaction range is somewhat arbitrary,
we check the validity of the kinetic approach by calculating $\lambda/r$ 
for all three estimates of $r$. 
The results agree qualitatively for a single scattering amplitude, 
but depend strongly on the scattering amplitude itself.
This is shown in fig.~\ref{fig:validity}, where
the different bands indicate the regions between the smallest and the 
largest $\lambda/r$ for a given scattering amplitude. 
For the weakest interaction, the temperature and momentum independent
Weinberg amplitude, the ratio $\lambda/r$ is much larger than unity 
for all temperatures of interest.
For the other approximations, on the other hand, there are regions where
$\lambda/r$ gets small and, thus, the Boltzmann approach is not valid.
In fact, since $\lambda$ is proportional to $\eta$ in our estimate,
these regions roughly coincide with the minima of the shear viscosity.
For instance, for the momentum independent, but temperature dependent 
NJL-model amplitude, $\lambda/r$ gets very small in the vicinity of 
$T_{diss}$ and $T_{Mott}$, whereas including the momentum dependence of 
the $\sigma$-propagator in RPA invalidates the BUU approach even at 
temperatures below 100~MeV.
This is cured again, when the $\pi\pi$-decay width of the 
$\sigma$-meson is taken into account. In this case the results can be
trusted until closely below the Mott temperature.

\subsection{Fluidity measures}

In order to compare the fluidity of systems with rather different
natural scales, like cold atomic gases, water or the quark-gluon 
plasma, the absolute value of the shear viscosity is not very meaningful.
More appropriate fluidity measures are therefore dimensionless ratios of 
$\eta$ and some thermal quantity, which scales in the proper way.
The most prominent example, at least in the context of the quark-gluon
plasma, is the ratio of shear viscosity and entropy density $s$,
which has been conjectured to have a universal lower bound,
$\eta/s \geq {1}/{4\pi}$~\cite{KSS:PRL05}.

An alternative measure, which is better suited for the comparison
of relativistic and nonrelativistic systems, has been suggested in 
ref.~\cite{LiaoKoch:PRC2010}.
It is given by the ratio of two characteristic length scales 
\beq
L_\eta / L_n = \frac{\eta n^{1/3}}{h c_s},
\eeq
where $L_\eta=\eta/(h c_s)$ is related to the shear viscosity and 
$L_n = 1/n^{1/3}$ is related to the density. 
Here $n$ is the number density, $c_s$ is the sound velocity and 
$h=\epsilon+p$ is the enthalpy density.

\begin{figure}
  \begin{center}
    \psfrag{LoL}{\hspace{-1ex}{\small{$L_\eta / L_n$}}}
    \psfrag{eos}{\hspace{-1ex}{\small{$\eta/s$}}}
    \psfrag{t}{\hspace{-3ex}$T [\MeV]$}
    \includegraphics[height=\columnwidth,angle=-90]{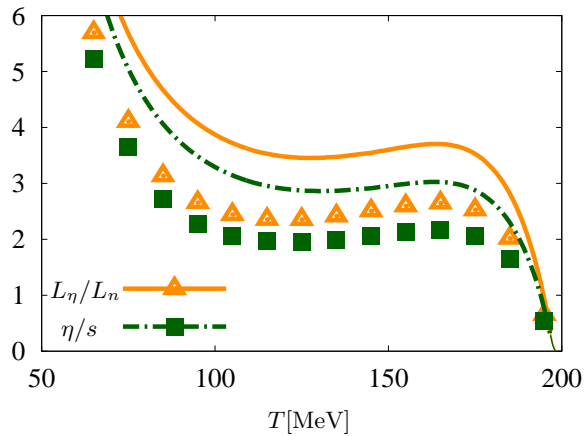}
    \caption{The ratios $\eta/s$ (green dash-dotted line) 
    and $L_\eta / L_n$ (orange solid line) as functions of temperature, 
    calculated with the dressed $\sigma$-meson. The simple estimate 
    $\eta \approx \bar{p}/(3\sigma(\bar{p}))$ is indicated by triangles for 
    $L_\eta / L_n$, and by squares for $\eta/s$. }
    \label{fig:eosLoL}
  \end{center}
\end{figure}

Our results for $\eta/s$ and $L_\eta / L_n$ are shown in 
fig.~\ref{fig:eosLoL}.
For the shear viscosity, we only take the most realistic amplitude, 
which includes the dressed $\sigma$-meson.
The other quantities involved in the ratios, {\it i.e.}, $s$, $n$, 
$h$ and $c_s$, 
are calculated in the ideal-gas limit, which is consistent with the 
Boltzmann approach.
We find that both fluidity measures are very similar. 
After decreasing by several orders of magnitude at low temperatures, 
the curves become rather flat in an intermediate temperature regime, 
where we find $\eta/s \approx 3$ and a somewhat larger value for 
$L_\eta / L_n$.
Finally, there is another steep decrease when $T$ approaches the Mott
temperature. However, as already mentioned, this drop is probably an 
artifact of neglecting the momentum dependence of the quark triangles 
and boxes.

For comparison we also show the ratios one obtains when
$\eta$ is calculated from the 
simple estimate eq.~(\ref{eq:etaest}) with the additional assumption that 
the thermally averaged cross section $\bar\sigma_{tot}$ can be approximated 
by the cross section  at the averaged momentum $\sigma_{tot}(\bar p)$.
Compared with the Boltzmann approach, the results are about 30\% lower but
in good qualitative agreement. Hence, at a stage where the uncertainties in 
the scattering amplitude are still very large, this approximation is a viable 
alternative to the exact solutions of the BUU equations.


\section{Discussion}
\label{sec:discussion}

In this article, we have computed the shear viscosity of a pion gas within
relativistic kinetic theory. Our main focus was to study the effects that 
originate from the chiral crossover transition. 
To this end we have employed a two-flavor NJL model to calculate the 
effective masses and scattering amplitudes of pions in a hot medium.
The shear viscosity was then obtained from the solution of the BUU
equation in a Chapman-Enskog expansion to leading order. 
Simultaneously we have carefully analyzed the range of validity of this 
approach, bearing in mind that kinetic theory can only be applied for dilute 
gases.

While at low temperatures our results are consistent with lowest-order
chiral perturbation theory, the scattering of pions is strongly affected by 
the restoration of chiral symmetry at high temperature. 
In particular the lowering of the $\sigma$-mass leads to a strong enhancement
of the $s$-channel $\sigma$-exchange diagram. As a consequence, the shear
viscosity is much smaller than it would be expected from a simple vacuum
extrapolation. 

In the cross-over region the $\pi\pi$-scattering length diverges at two 
characteristic temperatures, $T_{diss}$ and $T_{Mott}$. This is reminiscent 
of the situation in ultra-cold atomic gases in the vicinity of a Feshbach 
resonance. Motivated by this analogy, in a first step, we have neglected the 
momentum dependence of the scattering amplitude completely and only kept the 
temperature dependence according to the scattering length. Accordingly, the 
shear viscosity was found to have sharp minima at $T_{diss}$ and $T_{Mott}$ 
in this approximation. 

On the other hand, unlike for cold atoms, the knowledge of the scattering 
length is not sufficient to describe the scattering of hot pions.
Because of thermal motion, the $\sigma$-meson mass pole can be reached 
at temperatures well below the dissociation temperature, 
and we therefore find a considerable reduction of the shear viscosity 
at lower temperatures, when the momentum dependence of the scattering 
amplitude is taken into account. 

However, the results turned out to be very sensitive to the approximations 
we have applied to evaluate the $\sigma$-propagator, and our `most reliable'
approximation, where we have included the $\sigma \rightarrow \pi\pi$-decay
width, is surely not the last word. In fact, except for the region close
to the Mott temperature, where the Boltzmann approach
should not be trusted, our results for 
$\eta/s$ are still more than one order of magnitude above the conjectured
lower bound of $1/4\pi$. Thus, various improvements and extensions of the
model should be performed:

The dressing of the $\sigma$-meson as described in sec.~\ref{sec:Mpipi-dressed}
was done in a rather simple way by only including the imaginary part of the 
diagram shown in fig.~\ref{fig:sigma-pipi-loop}. As already pointed out,
this ensures the correct threshold behavior, dictated by chiral symmetry,
but it violates causality. 
One may try to cure this problem by calculating the real part of the 
of the diagram from a subtracted dispersion relation, with the 
subtraction constant chosen in such a way that the scattering lengths 
remain unchanged. This would be rather straightforward in the chiral
limit. For physical pion masses it is more difficult because,
even at threshold, the $\sigma$-meson in the $s$-channel is probed at a 
different kinematical point than in the $t$- and $u$-channel 
($q^2 = 4 m_\pi^2$ and $q^2 = 0$, respectively), 
so that it is not immediately clear how to make the subtraction. 

The dressed $\sigma$-meson can be viewed as to arise from iterating the 
$\sigma$-meson exchange diagram in fig.~\ref{fig:pipi-scattering-diagrams}.
In the same way one should also iterate the box diagram. 
Formally, this yields a series of diagrams which are of the same order
in $1/N_c$ as the dressing of the $\sigma$-meson and which contributes
to the imaginary part above the two-pion threshold as well. 
We have not considered these diagrams so far, because our primary interest
was in getting rid of the sharp resonance peak caused by the RPA 
$\sigma$-meson. They might nevertheless give important contributions 
to the scattering amplitude. 
Taking them into account, makes it of course even more difficult to
satisfy chiral theorems and causality, and one has to see whether 
dispersion techniques can help here as well. 
Maybe at that point one has to give up some of the formal requirements
and should judge the reliability of the approximation by comparison with 
$\pi\pi$-scattering data at $T=0$.

In the $\pi\pi$-sector, we should also include intermediate $\rho$-mesons in 
order to obtain a realistic description of the $p$-wave isovector channel.
In addition, we should include other hadrons, which are suppressed at low 
temperatures, but can become important in the crossover region
\cite{Muronga:2003tb,Gorenstein:2007mw,NoronhaHostler:2008ju}. In 
particular, we wish to extend the model to three flavors and include kaons
and $\eta$-mesons. Moreover, we would like to include the scattering of
quarks, which become important above the crossover temperature.  
On the other hand, in order to suppress quark effects in the hadronic phase,
they should be coupled to the Polyakov loop. Work in these directions is
in progress.

\begin{acknowledgments}
This work was supported in part by the Helmholtz International Center for FAIR, the 
Helmholtz Institute EMMI and the BMBF grant 06DA9047I. 
\end{acknowledgments}

\appendix


\section{Generalized Sonine functions}
\label{sec:sonine}

In order to solve the linear integral equation (\ref{eq:BUUlin}), we
follow Refs.~\cite{DLlE:04,IOM:08} and expand the scalar function 
${\cal{B}}(p)$ defined in eq.~(\ref{eq:Bp}) in so-called generalized 
Sonine polynomials $P^r_p \equiv P^r(|\vec p|)$,
\beq
\label{eq:B-Sonine-expansion}
{\cal{B}}(|\vec p|) = \sum_r b^r P^r_p.
\eeq
The latter are polynomials of degree $r$,
\beq
P^r_p = \sum_{j=0}^{r} c_j^r\,p^j,
\eeq
which satisfy the orthogonality relations 
\beq
\label{eq:sonine-orthogonality}
\int \intd p \, \frac{p^4}{E_p}f^{(0)}_p \left((1+f^{(0)}_p\right)) 
P^r_p P^s_p 
= {\mathcal{N}}^{r}\delta^{rs}.
\eeq
As in Refs.~\cite{DLlE:04,IOM:08} we do not choose the normalization 
constants ${\mathcal{N}}^{r}$ to be unity, but consider monic functions,
$c_r^r=1$. The remaining coefficients $c_j^r$ are then uniquely determined
by eq.~\eqref{eq:sonine-orthogonality}. These coefficients depend on
$T$, $\mu_\pi$, and $m_\pi$, but the numerical orthogonalization is not a 
challenge.

Thus, the remaining problem is to compute the coefficients $b^r$ 
of the expansion (\ref{eq:B-Sonine-expansion}).
To this end, we multiply  both sides of eq.~(\ref{eq:BUUlin}) with
\beq
P^{rij}({\vec p}) =
P^r_p \left( \hat{p}^i \hat{p}^j -\frac{1}{3}\delta^{ij} \right),
\eeq
sum over the indices $i$ and $j$, and integrate over the momentum $\vec{p}$.
Then, after inserting eq.~(\ref{eq:Bp}) and the expansion 
(\ref{eq:B-Sonine-expansion}) for $B^{ij}$, one obtains a linear equation for
the expansion coefficients,
\beq
\label{eq:rkt-abc}
A^{rs} b^s = C^r,
\eeq
with
\begin{align}
&A^{rs} 
\;=\;
\nonumber \\  
&
\frac{g_\pi}{2}
  \intddd{p} \intddd{p'} \intddd{p_1} \intddd{p_1'} \,\times
\nonumber \\  
 &  \hspace{5mm} \times 
  (2\pi)^4\delta^4(p+p_1-p'-p_1') \, 
  \frac{|{\cal{M}}_{\pi \pi}|^2 }{16EE_1E'E_1'}\, \times
\nonumber \\ 
&  \hspace{5mm} \times
  f^{(0)}_p f^{(0)}_{p_1}
  (1+f^{(0)}_{p'})(1+ f^{(0)}_{p_1'}) \, 
  P^r_p \left( \hat{p}^i \hat{p}^j -\frac{1}{3}\delta^{ij} \right) \, \times 
\nonumber \\ 
& \hspace{5mm} \times
  \left[ P^s_{p'} \left( \hat{p}'\,^i \hat{p}'\,^j -\frac{1}{3}\delta^{ij} \right)
+ P^s_{p_1'} \left( \hat{p}_1'\,^i \hat{p}_1'\,^j -\frac{1}{3}\delta^{ij} \right) 
\right.
\nonumber \\ 
& \hspace{7mm}
\left.
- P^s_p  \left( \hat{p}^i \hat{p}^j -\frac{1}{3}\delta^{ij} \right)
- P^s_{p_1}  \left( \hat{p}_1^i \hat{p}_1^j -\frac{1}{3}\delta^{ij} \right)
\right],
\end{align}
and
\beq
C^{r} 
\;=\;
-\frac{2}{3}\intddd{p} \frac{\vec p^2}{ET}\,f^{(0)}_p (1+f^{(0)}_{p})\, P^r_p .
\eeq
The tensor $A^{rs}$ and the vector $C^r$ have to be calculated
numerically. After that the coefficients $b_r$ are obtained by
inverting eq.~(\ref{eq:rkt-abc}).

\begin{figure}
  \begin{center}
    \psfrag{eta}{\hspace{-6ex}{{$\eta[N_S]/\eta[N_S=5]$}}}
    \psfrag{Ns}{{{\hspace{-3ex}$N_S$}}}
    \psfrag{mfp}{\hspace{0ex}{\small{mfp}}}
    \psfrag{dressed}{\hspace{-2.5ex}{\small{dressed}}}
    \psfrag{RPA}{\hspace{-2.5ex}{\small{RPA}}}
    \psfrag{Weinberg}{\hspace{-2.5ex}{\small{Weinberg}}}
    \includegraphics[height=\columnwidth,angle=-90]{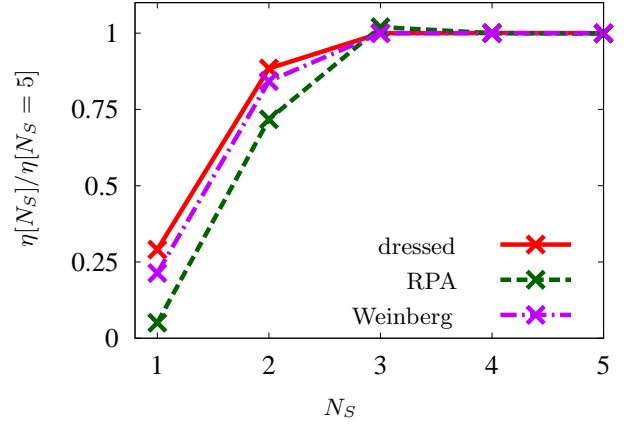}
    \caption{Shear viscosity $\eta$ at $T=170\,\MeV$ obtained with
different numbers $N_S$ of generalized Sonine functions, normalized to the
corresponding value for $N_S=5$. The points have been connected by lines to
guide the eye. The different colors and line styles indicate different 
approximations for the scattering amplitude and are consistent with the
choice in fig.~\ref{fig:eta}. The result for the temperature dependent but
momentum independent amplitude is identical to the result for the
Weinberg amplitude, since the amplitudes differ only by a temperature
dependent factor, which drops out by the normalization.}
    \label{fig:Sonine}
  \end{center}
\end{figure}

For the numerical treatment, the expansion has to be truncated to a finite 
set of generalized Sonine polynomials.
We have checked the convergence by studying the dependence of the
resulting shear viscosity on the number of basis functions $N_S$.
An example is given in fig.~\ref{fig:Sonine}, where the $N_S$ dependence of
$\eta$ is shown for $T=170\,\MeV$.
We find that it is sufficient to take into account the first three 
polynomials, a result which holds in the whole temperature range 
$T<T_{Mott}$. 

\newpage


\end{document}